\documentclass[structabstract]{aa}
\usepackage{color}
\usepackage{graphicx}
\usepackage{txfonts}
\usepackage{lmodern}
\usepackage{natbib,twoopt}
\usepackage[breaklinks=true]{hyperref}
\usepackage{todonotes}
\usepackage{lscape}
\usepackage{multirow}

\hypersetup{colorlinks=true,linkcolor=blue,citecolor=blue,urlcolor=blue}

\definecolor{red}{rgb}{1,0,0}
\definecolor{blue}{rgb}{0,0,1}
\definecolor{mygreen}{RGB}{0,128,0}

\graphicspath{{pic/}}

\bibpunct{(}{)}{;}{a}{}{,} 
\makeatletter
\newcommandtwoopt{\citeads}[3][][]{\href{https://ui.adsabs.harvard.edu/\#abs/#3}%
{\def\hyper@linkstart##1##2{}%
\let\hyper@linkend\@empty\citealp[#1][#2]{#3}}}
\newcommandtwoopt{\citepads}[3][][]{\href{https://ui.adsabs.harvard.edu/\#abs/#3}%
{\def\hyper@linkstart##1##2{}%
\let\hyper@linkend\@empty\citep[#1][#2]{#3}}}
\newcommandtwoopt{\citetads}[3][][]{\href{https://ui.adsabs.harvard.edu/\#abs/#3}%
{\def\hyper@linkstart##1##2{}%
\let\hyper@linkend\@empty\citet[#1][#2]{#3}}}
\newcommandtwoopt{\citeyearads}[3][][]%
{\href{https://ui.adsabs.harvard.edu/\#abs/#3}
{\def\hyper@linkstart##1##2{}%
\let\hyper@linkend\@empty\citeyear[#1][#2]{#3}}}
\makeatother

\begin{document}
        
\title{The convective surface of the red supergiant Antares\thanks{Based on observations made with ESO telescopes at Paranal Observatory, under ESO programs 093.D-0378(A), 093.D-0378(B), 093.D-0378(C) and 093.D-0673(C).}}
\subtitle{VLTI/PIONIER interferometry in the near infrared}
\titlerunning{The convective surface of Antares}

\author{
        M.~Montarg\`es\inst{1}
        \and
        A.~Chiavassa\inst{2}
        \and
        P.~Kervella\inst{3,4}
        \and
        S.~T.~Ridgway\inst{5}
        \and
        G.~Perrin\inst{4}
        \and
        J.-B.~Le~Bouquin\inst{6}
        \and
        S.~Lacour\inst{4}
}

\institute{
        Institut de Radioastronomie Millim\'etrique, 300 rue de la Piscine, 38406, Saint Martin d'H\`eres, France
        \and
        Universit\'e C\^ote d’Azur, Observatoire de la C\^ote d’Azur, CNRS, 
        Lagrange, CS 34229, Nice, France
        \and
        Unidad Mixta Internacional Franco-Chilena de Astronom\'{i}a (UMI 3386), CNRS/INSU, France
        \& Departamento de Astronom\'{i}a, Universidad de Chile, Camino El Observatorio 1515, Las Condes, Santiago, Chile.
        \and
        LESIA, Observatoire de Paris, PSL Research University, CNRS UMR 8109, Sorbonne Universit\'es, UPMC, Universit\'e Paris Diderot, Sorbonne Paris Cit\'e,
        5 place Jules Janssen, F-92195 Meudon, France
        \and
        National Optical Astronomy Observatories, 950 North Cherry Avenue, Tucson, AZ 85719, USA
        \and
        UJF-Grenoble 1/CNRS-INSU, Institut de Plan\'etologie et d'Astrophysique de 
        Grenoble (IPAG), UMR 5274, Grenoble, France     
}

\date{Received 31 Oct. 2016; Accepted 15 May 2017}

\abstract
{Convection is a candidate to explain the trigger of red supergiant star (RSG) mass loss. Owing to the small size of the convective cells on the photosphere, few of the characteristics of RSGs are known.}
{Using near infrared interferometry, we intend to resolve the photosphere of RSGs and to bring new constraints on their modeling.}
{We observed the nearby red supergiant Antares using the four-telescope instrument 
VLTI/PIONIER. We collected data on the three available configurations of the 
1.8m telescopes in the H band.}
{We obtained unprecedented angular resolution on the disk of a star (6\% of the 
star angular diameter) that limits the mean size of convective cells and offers 
new constraints on numerical simulations. Using an analytical model with a distribution of bright 
spots we determine their effect on the visibility signal.}
{We determine that the interferometric signal on Antares is compatible with convective cells of various sizes from 45\% to 5\% of the angular diameter. We also conclude that convective cells can strongly affect the angular diameter and limb-darkening measurements. In particular, the apparent angular diameter becomes dependent on the sampled position angles.}

\keywords{Stars: individual: Antares; Stars: imaging; Stars: supergiants; Stars: mass-loss; Infrared: Stars; Techniques: interferometric}

\maketitle

\section{Introduction}

Chemical enrichment of the Universe is driven by evolved stars. Although currently rare, massive stars were much more numerous during the early times of 
the Universe. When entering the red supergiant (RSG) phase in their later 
evolution, these stars experience intense mass loss. The material expelled from 
the star cools, allowing the formation of molecules and dust that will be 
essential contributions to new stellar systems. However, the processes driving 
this outflow of material remain only partly understood.

RSG stars do not experience flares or large-scale pulsations that could inject 
enough momentum for the material to be launched away from the star. 
\citetads{2015A&A...575A..50A} demonstrated that pulsation models do not reproduce the molecular extension of the atmosphere of RSG that would be consistent with their mass loss. \citetads{2010ASPC..425..152H} stated that 
there is no available physical scenario that can be shown to initiate the 
outflow of massive evolved stars.

\citetads{1975ApJ...195..137S} predicted that, compared to the sun, the photosphere of RSG could be host to a smaller number of much larger convective cells. From spectroscopic observations of a sample of RSGs, \citetads{2007A&A...469..671J} proposed that large convective cells could trigger mass loss by locally lowering the effective gravity and allowing radiative pressure on molecular lines to initiate the outflow. 

Early imaging observations offered further evidence for convective activity. 
Only two large hot spots were detected over the visible hemisphere of 
\object{Betelgeuse} by \citetads{2009A&A...508..923H} in a high-dynamic-range 
reconstructed image from IOTA H band interferometric observations. 
\citetads[][hereafter C10b]{2010A&A...515A..12C} identified a possible 
convective pattern in the same dataset using 3D radiative hydrodynamics 
simulations (RHD).\\

\object{Antares} ($\alpha$ Sco, HD 148478, HR 6134) is the closest RSG ($\pi = 5.89 \pm 1.00$~mas, \citeads{2007A&A...474..653V}). Its apparent diameter of $\sim 37$~mas, measured from VLTI/AMBER data by \citetads[][hereafter O13]{2013A&A...555A..24O}, makes it one of the largest stars in our sky. The same authors derived a mass of $15 \pm 5$~M$_\odot$ and an age of 11-15~Myr. It is a classical RSG with a spectral type M0.5Iab. It has a B3V companion (\object{Antares B}) located far enough away (2.7", \citeads{2014A&A...568A..17O}) that we can consider the primary alone in the present study.

Around this RSG, O13 observed upward and downward motions of CO in the upper atmosphere or near circumstellar region (1.3~R$_\star$), strongly pointing towards a convection-based mechanism. Observations of the circumstellar environment of Antares by \citetads{2014A&A...568A..17O} using the VLT/VISIR instrument revealed a clumpy and dusty envelope. This is compatible with convection-triggered mass loss.
        
\citetads{2013ApJ...777...10P} studied the short timescale radial velocity variations of the star. In addition to large hot spots, reported earlier on RSG, they suggested that this observation could be associated to smaller-scale convective activity. \citetads{2010ApJ...725.1170S} showed that the long secondary photometric period of RSGs, and in particular of Antares, could be related to convection. Therefore we see that there are several reasons for suspecting the importance of convection in RSG atmospheres, and for hypothesizing that it is triggering the mass loss.  However, an observational basis for constraining the spatial scale of convection is needed for empirical and theoretical modeling of the mass loss.\\

Until now, near-infrared interferometric observations of RSGs have only probed the spatial frequencies up to the fifth or sixth lobe of the visibility function at most. Convective simulations from \citetads[][hereafter C11a]{2011A&A...535A..22C} predicted that the convective cell structures could enhance the visibility signal up to the tenth lobe at least. The smaller granules detected at those high spatial frequencies are expected to be more numerous. They have yet to be detected with optical interferometry. This is essential to minimally constrain the surface convection pattern on RSGs.\\

Interferometry is the only way to obtain detailed observations of the photospheric region of RSGs. We present VLTI/PIONIER observations of Antares at a very high angular resolution ($\sim 1/15^\mathrm{th}$ of the stellar radius). In Sect. \ref{Sect:Observations} we present our observations and the data reduction. We fit the data with analytical models in Sect. \ref{Sect:Analytical}, ranging from classical disks to a distribution of models that includes bright Gaussian spots. We continue the analysis with 3D radiative hydrodynamics simulations in Sect. \ref{Sect:RHD_Sim}.

\section{Observations\label{Sect:Observations}}

\begin{figure}
        \centering
        \resizebox{\hsize}{!}{\includegraphics{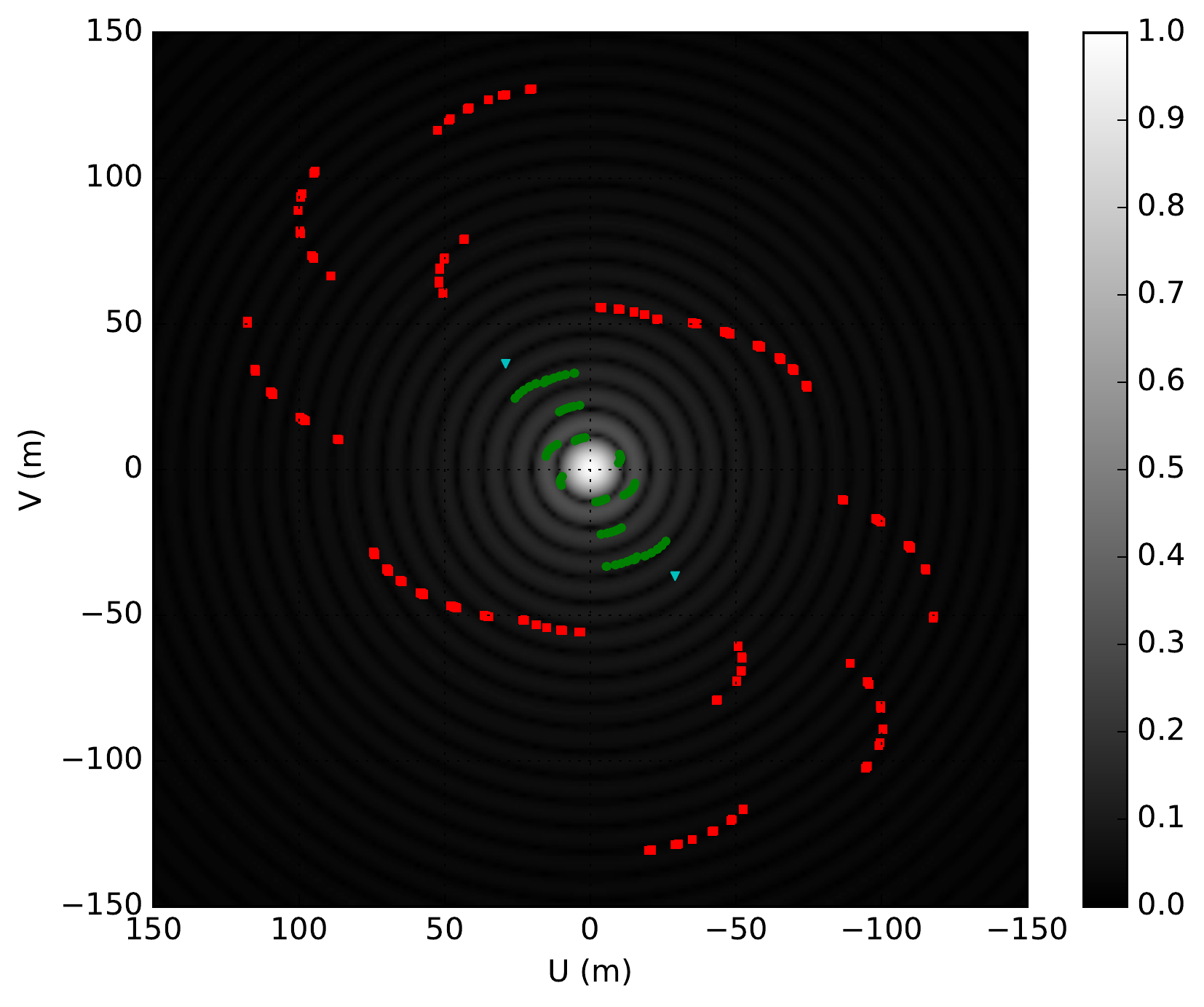}}
        \caption{($u, v$) coverage of our VLTI/PIONIER observations of Antares. North is up and East is left. The compact AT configuration is represented with the green circles, the medium configuration with the blue triangles and the extended configuration is represented with the red squares. Underneath, the visibility amplitude of a power-law limb-darkened disk matching the best fit parameters for Antares at 1.61~$\mu$m is represented (see Sect \ref{Sect:Analytical}).\label{Fig:uvcov}}
\end{figure}

To observe Antares, we used the European Southern Observatory's Very Large Telescope Interferometer (VLTI, \citeads{2010SPIE.7734E...3H}) located on top of Cerro Paranal in Northern Chile with the PIONIER instrument (Precision Integrated-Optics Near-infrared Imaging ExpeRiment, \citeads{2011A&A...535A..67L}), which recombines the light of four telescopes simultaneously. We observed with the four 1.8~m diameter Auxiliary Telescopes (AT) of the VLTI in three different configurations that gave us access to baselines from 11.3~m to 153.0~m on the ground. The resulting ($u, \,v$) coverage is represented in Fig. \ref{Fig:uvcov}. The instrument was configured in the high spectral resolution mode (R $\sim 40$) that produces seven spectral channels over the H band ($1.54 - 1.80~\mu$m). To avoid saturation, without using a neutral density, we read only the three central pixels of the detector ($1.60 - 1.71~\mu$m). Antares and its calibrators were observed on 2014 April 24, 29 and May 4 and 7.

The data were reduced using the publicly available PIONIER pipeline \citepads{2011A&A...535A..67L}. We adopted the angular diameters of Table \ref{Tab:calibrators_data} for the calibrators, obtained using the JMMC tool \texttt{SearchCal}\footnote{Available at \url{http://www.jmmc.fr/searchcal}} \citepads{2006A&A...456..789B,2011A&A...535A..53B}. The pipeline automatically computes the uncertainties: on the uncalibrated data it derives the statistical dispersion over 100 scans each of $\sim$ 30~s exposure. Then for the calibrated product it quadratically adds the error from the transfer function. As PIONIER is a four-telescope instrument, we finally got six squared visibility and four closure phase measurements per observation and per spectral channel. We had to ignore the data from baseline A1-K0 because of their poor signal to noise ratio.

\begin{table}
        \caption{Adopted uniform disk diameters for the interferometric calibrators. (\citeads{2006A&A...456..789B,2011A&A...535A..53B} and \citeads{2005A&A...433.1155M})}
        \label{Tab:calibrators_data}
        \centering
        \begin{tabular}{ll}
                \hline \hline
                \noalign{\smallskip}
                Name & Diameter \\
                 & (mas) \\
                \hline
                \noalign{\smallskip}
                \object{HR 5969} & $ 1.79 \pm 0.13$ \\
                \object{HR 6145} & $0.84 \pm 0.06$ \\   
                \object{HD 142407} & $1.27 \pm 0.09$ \\
                \object{HD 143900} & $1.28 \pm 0.05$ \\
                \object{HD 148643} & $1.47 \pm 0.08$ \\
                \object{$\psi$ Oph} & $1.89 \pm 0.13$ \\
                \hline
        \end{tabular}
\end{table}

\section{Analytical model analysis\label{Sect:Analytical}}

\subsection{Classical disk models \label{Sect:UD_LDD}}

To derive the angular diameter of the star, we use two different models 
that are commonly employed in the literature for RSGs: a uniform disk (UD) and 
a power law limb-darkened disk (LDD, with $I(\mu)/I_0 = \mu^\delta$). The 
squared visibility of the later is given by \citetads{1997A&A...327..199H} :

\begin{equation}
V_\mathrm{LDD}(u,v) = \Gamma(\nu+1)\frac{J_\nu (x)}{(x/2)^\nu}\label{Eq:VisLDD}
,\end{equation}

where $\nu = \delta/2 +1$ and $\Gamma$ is the Euler function.

No bandwidth smearing was used in any of the following fits as tests showed that its effect kept the derived values within their error bars.

We first fitted the entire squared visibility dataset (all baseline lengths and spectral channels). However we noticed important positive deviations from the simple disk models indicating the presence of small additional features on the stellar surface. The reduced $\chi^2$ ($\tilde{\chi}^2$ hereafter) reaches $\sim 180$ for the UD model and $\sim 25$ for the LDD.
        
To avoid the contamination of possible small scale features (also suggested by the closure phase deviations from 0$^\circ$ or 180$^\circ$ on Fig. \ref{Fig:Nspot_best}), we then only fitted the first lobe of the squared visibility function (spatial frequencies lower than 35~arcsec$^{-1}$) for the UD model and the first two lobes for the LDD model (spatial frequencies lower than 50~arcsec$^{-1}$). The $\tilde{\chi}^2$ is  $\sim 20$ and $\sim 16$, respectively, meaning that important deviations from the models are present.
 
\begin{figure}
        \centering
        \resizebox{\hsize}{!}{\includegraphics{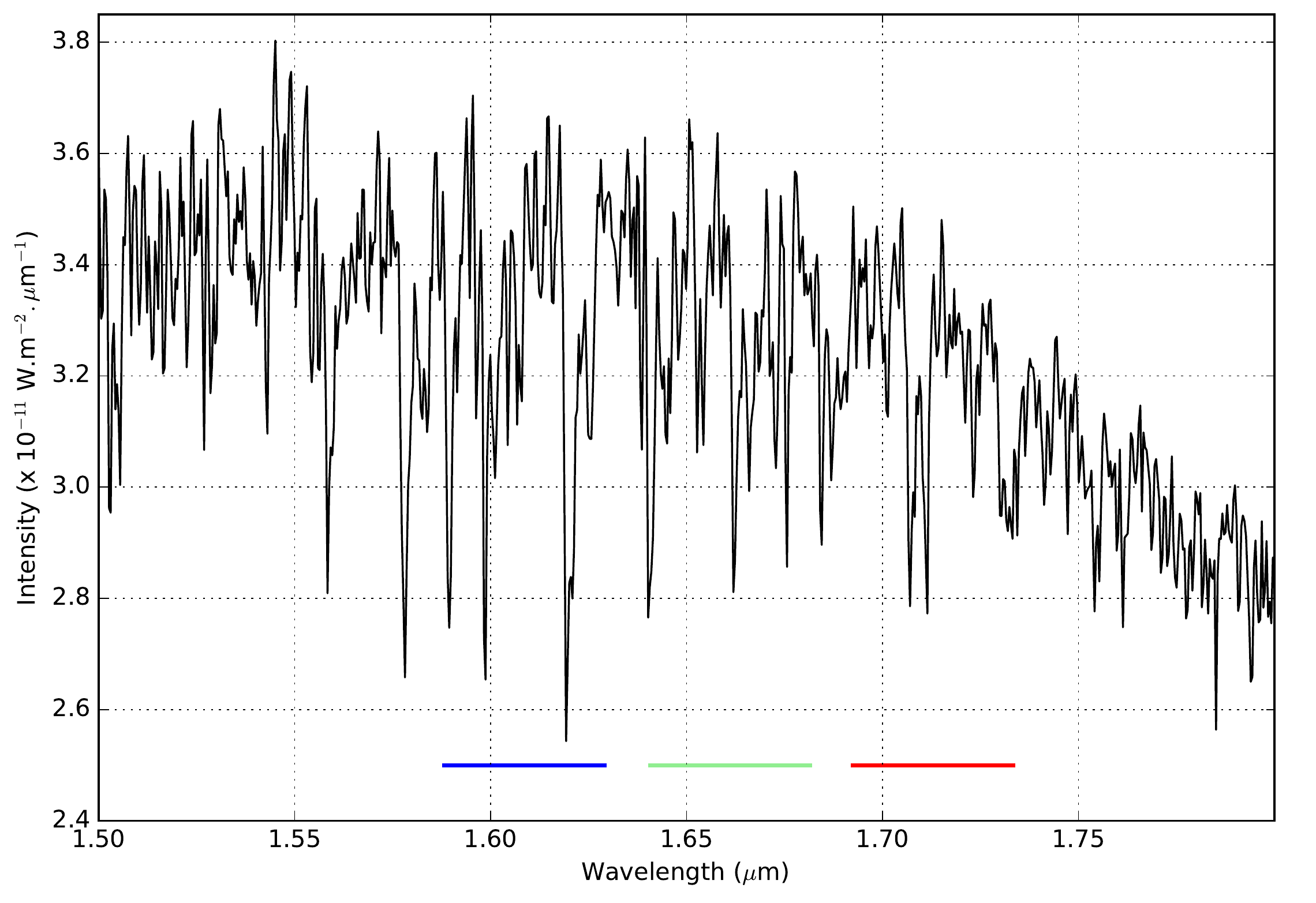}}
        \caption{Spectrum of the M0Ib-II star \object{HD236697}, in the H band, from the Infrared Telescope Facility spectral library \citepads{2009ApJS..185..289R}. The PIONIER spectral channels are indicated with colored horizontal lines. \label{Fig:spect_Ant}}
\end{figure}

Our VLTI/PIONIER observations consist of three distinct spectral channels probing various molecular lines (Fig. \ref{Fig:spect_Ant}). Fitting those three channels separately in the first and first two lobes for the UD and LDD models,  respectively, lowers the $\tilde{\chi}^2$ to 18-56 for the UD and  7-10 for the LDD. Therefore, we can conclude that the star does not look the same in these three different wavelengths in the H band. Still, some deviations cannot be reproduced by the models.

\begin{figure}
        \centering
        \resizebox{\hsize}{!}{\includegraphics{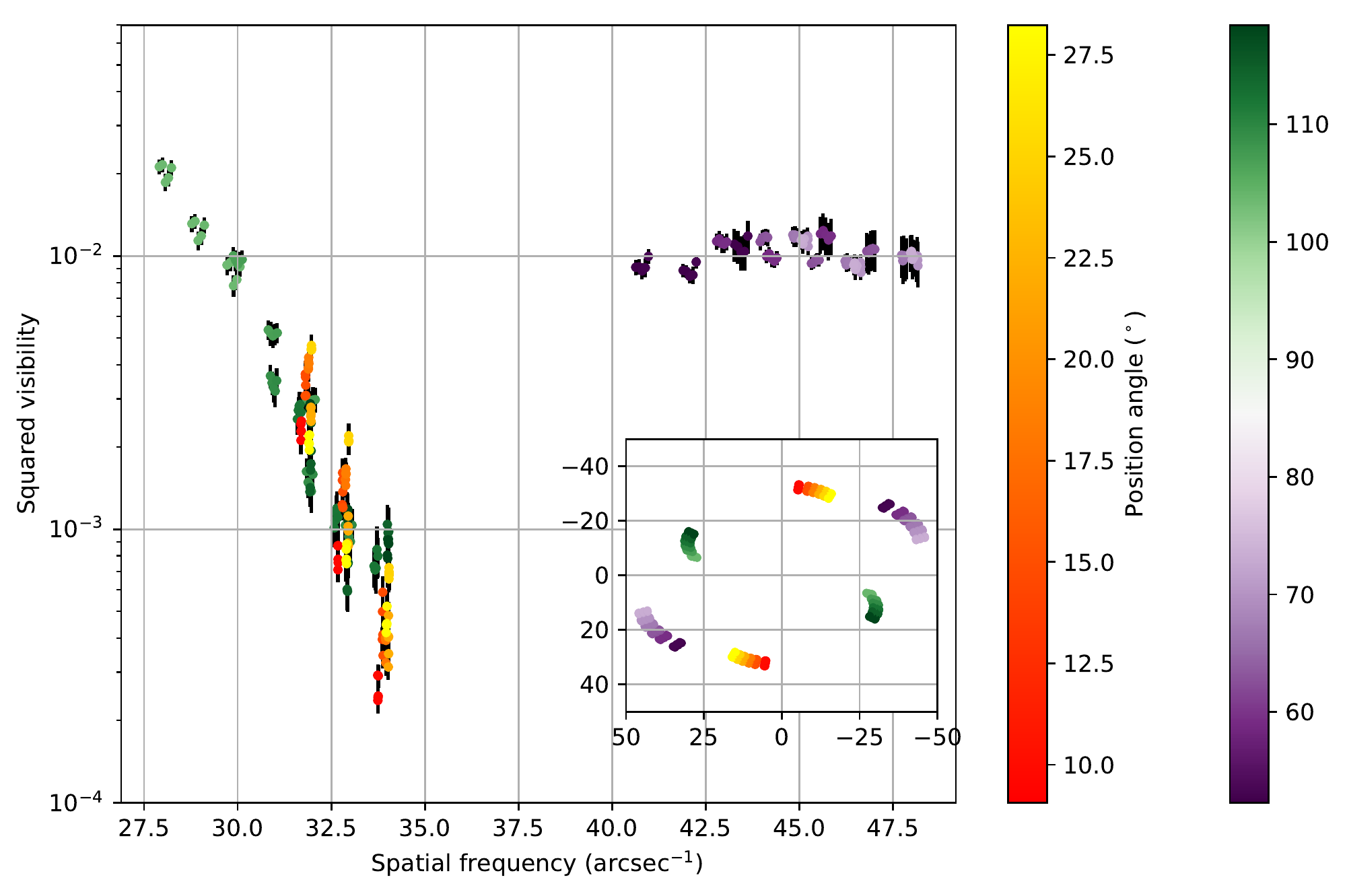}}
        \caption{VLTI/PIONIER squared visibilities measured on Antares as a function of spatial frequency for the first two lobes. In the inset, the ($u ,v $) coverage is represented in arcsec$^{-1}$. In both sub-figures, the PA is color-coded.\label{Fig:uv_v2_azi}}
\end{figure} 

\begin{figure*}
        \resizebox{\hsize}{!}{\includegraphics{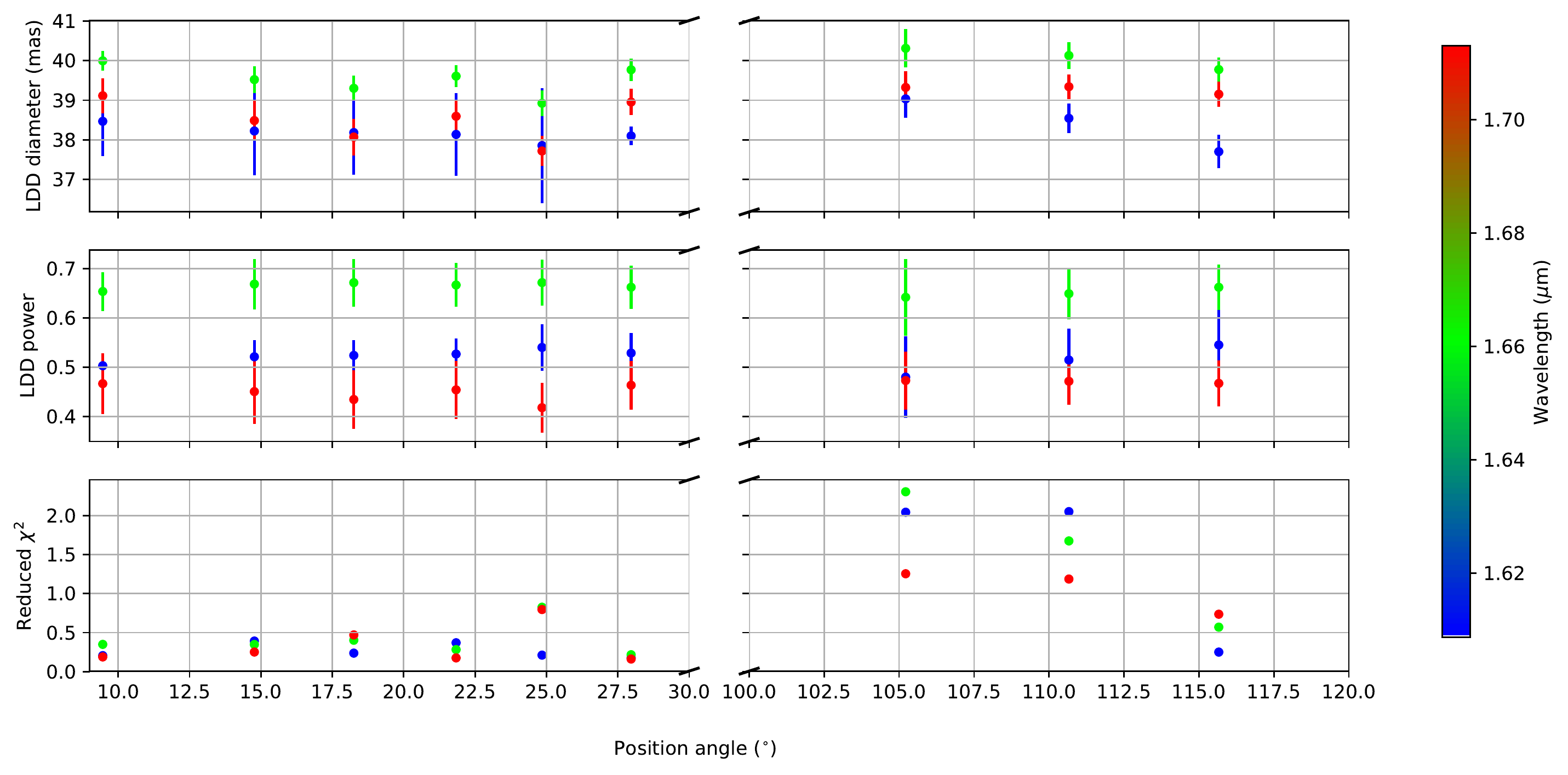}}
        \caption{LDD best fit parameters with a PA and spectral channel selection of the squared visibilities in the first and second lobes (spatial frequencies below 50 arcsec$^{-1}$). \label{Fig:fit_LDD_az}}
\end{figure*}

In addition to the spectral channels, our observations cover several position angle (PA) directions in the first and second lobes. Figure \ref{Fig:uv_v2_azi} represents the PA dependency of the squared visibilities in the first two lobes. Three areas can be distinguished. In the first lobe, the baseline A1-B2 (green in Fig. \ref{Fig:uv_v2_azi}) probes baseline lengths between 9.8 and 11.4m and PA between 103$^\circ$ and 119$^\circ$. The squared visibility mainly depends on the spatial frequency and depends weakly on the PA. The same can be said of baseline A1-C2 (violet in Fig. \ref{Fig:uv_v2_azi}, baseline lengths in [14.3; 16 m], PA in [50$^\circ$; 80$^\circ$]). For baseline B2-C1 (yellow-orange in Fig. \ref{Fig:uv_v2_azi}), the situation is different: over a small baseline length range (11.1 to 11.4 m) and a significant PA range (9$^\circ$ to 29$^\circ$), the squared visibilities show great variation. As we do not have more azimuthal information, we made the following choice to build sub-samples: considering the fast squared visibility variation in PA for baseline B2-C1, each record was put in a different sub-sample. For the A1-B2 baseline and its slow squared visibility variation with PA, we defined three sub-samples equally spread in PA. Finally, the second lobe data (A1-C2) were used in every sample as they are dominated by uv-radius variation. We fitted a UD and LDD power law model for each spectral channel and each PA sub-sample. The results are presented in Table \ref{Tab:UD_LDD_param}. For the LDD, the best fit values are summarized in Fig. \ref{Fig:fit_LDD_az}.

\begin{table*}
        \caption{UD and LDD best fit parameters with a PA and spectral channel selection of the squared visibilities in the first and second lobes (spatial frequencies below 50 arcsec$^{-1}$). The given PA is associated to the first lobe data.}
        \label{Tab:UD_LDD_param}
        \centering
        \begin{tabular}{lllllll}
                \hline\hline\noalign{\smallskip}
                PA range ($^{\circ}$) and baselines & Wavelength ($\mu$m) & $\theta_\mathrm{UD}$ (mas) &  $\tilde{\chi}^2_\mathrm{UD}$ & $\theta_\mathrm{LDD}$ (mas) & $\delta_\mathrm{LDD}$ & $\tilde{\chi}^2_\mathrm{LDD}$ \\
                \hline\noalign{\smallskip}
114.48 - 117.95 &             1.609 &  $34.54 \pm 2.32$ &    13.0 &  $37.70 \pm 0.42$ &  $0.55 \pm 0.07$ &      0.3 \\
A1-B2 + A1-C2 &             1.661 &  $35.62 \pm 3.17$ &   146.9 &  $39.77 \pm 0.30$ &  $0.66 \pm 0.05$ &      0.6 \\
&             1.713 &  $36.09 \pm 1.47$ &    39.4 &  $39.15 \pm 0.32$ &  $0.47 \pm 0.05$ &      0.7 \\
109.01 - 112.34 &             1.609 &  $35.56 \pm 1.92$ &    13.7 &  $38.54 \pm 0.37$ &  $0.51 \pm 0.06$ &      2.1 \\
A1-B2 + A1-C2 &             1.661 &  $36.38 \pm 1.05$ &   139.3 &  $40.13 \pm 0.34$ &  $0.65 \pm 0.05$ &      1.7 \\
&             1.713 &  $36.31 \pm 1.68$ &    38.9 &  $39.34 \pm 0.31$ &  $0.47 \pm 0.05$ &      1.2 \\
103.35 - 107.10 &             1.609 &  $36.43 \pm 1.86$ &    12.2 &  $39.03 \pm 0.47$ &  $0.48 \pm 0.08$ &      2.0 \\
A1-B2 + A1-C2 &             1.661 &  $42.13 \pm 3.22$ &    57.0 &  $40.31 \pm 0.48$ &  $0.64 \pm 0.08$ &      2.3 \\
&             1.713 &  $36.24 \pm 2.81$ &    39.4 &  $39.32 \pm 0.41$ &  $0.47 \pm 0.06$ &      1.3 \\
9.07 - 9.85     &             1.609 &  $35.45 \pm 1.04$ &    13.3 &  $38.47 \pm 0.87$ &  $0.50 \pm 0.02$ &      0.2 \\
B2-C1 + A1-C2 &             1.661 &  $36.03 \pm 3.06$ &   157.7 &  $39.99 \pm 0.25$ &  $0.65 \pm 0.04$ &      0.3 \\
&             1.713 &  $35.96 \pm 1.67$ &    40.3 &  $39.11 \pm 0.44$ &  $0.47 \pm 0.06$ &      0.2 \\
14.40 - 15.14   &             1.609 &  $35.14 \pm 1.37$ &    13.8 &  $38.22 \pm 1.11$ &  $0.52 \pm 0.03$ &      0.4 \\
B2-C1 + A1-C2 &             1.661 &  $35.23 \pm 1.41$ &   150.4 &  $39.52 \pm 0.34$ &  $0.67 \pm 0.05$ &      0.3 \\
&             1.713 &  $35.25 \pm 1.95$ &    32.2 &  $38.49 \pm 0.51$ &  $0.45 \pm 0.07$ &      0.3 \\
17.90 - 18.60   &             1.609 &  $35.08 \pm 1.28$ &    13.7 &  $38.18 \pm 1.06$ &  $0.52 \pm 0.03$ &      0.2 \\
B2-C1 + A1-C2 &             1.661 &  $35.00 \pm 1.48$ &   145.6 &  $39.30 \pm 0.32$ &  $0.67 \pm 0.05$ &      0.4 \\
&             1.713 &  $34.96 \pm 1.51$ &    27.9 &  $38.07 \pm 0.46$ &  $0.43 \pm 0.06$ &      0.5 \\
21.50 - 22.16   &             1.609 &  $35.02 \pm 1.27$ &    13.9 &  $38.14 \pm 1.04$ &  $0.53 \pm 0.03$ &      0.4 \\
B2-C1 + A1-C2 &             1.661 &  $35.45 \pm 3.13$ &   152.9 &  $39.61 \pm 0.28$ &  $0.67 \pm 0.04$ &      0.3 \\
&             1.713 &  $35.53 \pm 1.74$ &    34.9 &  $38.59 \pm 0.41$ &  $0.45 \pm 0.06$ &      0.2 \\
24.54 - 25.15   &             1.609 &  $34.71 \pm 1.80$ &    13.8 &  $37.86 \pm 1.45$ &  $0.54 \pm 0.05$ &      0.2 \\
B2-C1 + A1-C2 &             1.661 &  $34.41 \pm 1.73$ &   131.9 &  $38.92 \pm 0.32$ &  $0.67 \pm 0.05$ &      0.8 \\
&             1.713 &  $34.76 \pm 2.97$ &    24.7 &  $37.72 \pm 0.38$ &  $0.42 \pm 0.05$ &      0.8 \\
27.69 - 28.24   &             1.609 &  $34.99 \pm 1.47$ &    13.7 &  $38.10 \pm 0.24$ &  $0.53 \pm 0.04$ &      0.2 \\
B2-C1 + A1-C2 &             1.661 &  $35.68 \pm 3.12$ &   155.4 &  $39.77 \pm 0.28$ &  $0.66 \pm 0.04$ &      0.2 \\
&             1.713 &  $35.98 \pm 1.52$ &    39.6 &  $38.95 \pm 0.33$ &  $0.46 \pm 0.05$ &      0.2 \\
                \hline
        \end{tabular}
        \tablefoot{The standard deviations are derived by determining the values of the parameters that give $\tilde{\chi}^2 = 2 \tilde{\chi}^2\mathrm{min}$. For the over-fitted cases ($\tilde{\chi}^2$<1), we used $\tilde{\chi}^2 = 2 $ to account for the higher contribution of the data noise this implies.}
\end{table*}

The UD model still poorly reproduces the data. However, with this selection of sub-datasets, the $\tilde{\chi}^2$ for the LDD model has values very close to 1 (and often below 1). This indicates that when separated according to their wavelength and PA, the data can be reproduced by a LDD model. The minimum angular diameter is, most of the time, reached for the shorter wavelength, which is consistent with this spectral channel being closest to the minimum H$^-$ opacity \citepads[][p155]{2008oasp.book.....G}. Additionally, we remark that the LDD power-law coefficient remains relatively constant compared to the angular diameter variations. This is probably caused by the presence of the same second lobe data in each sub-dataset: according to \citet{1974MNRAS.167..475H}, the second lobe is more sensitive to the LD effect than the first lobe. Unfortunately, we do not have other PA data points to perform this fit differently.
        
Although it is possible to use different disk models for different spectral channels, it is against the principle of a disk model to discriminate along the PA. We adopt the weighted and averaged-over-PA parameters from Table \ref{Tab:LDD_by_wlen} for the LDD model in each spectral channel. 

C11a and \citetads{2016A&A...588A.130M} showed that PA discrepancies in the LDD modeling of RSG were compatible with feature(s) on the stellar photosphere. Therefore, we favor this hypothesis and analyze our data with models capable of reproducing this effect. In particular, we do not use 1D wavelength dependent spherical models as they are not able to reproduce these PA variations.

\begin{table}
        \caption{Mean LDD parameters derived in each spectral channel from the PA-dependent fit (Table \ref{Tab:UD_LDD_param}). \label{Tab:LDD_by_wlen}}
        \centering
        \begin{tabular}{llll}
                \hline\hline\noalign{\smallskip}
                Parameter & 1.61 $\mu$m & 1.66 $\mu$m & 1.71 $\mu$m \\
                \hline\noalign{\smallskip}
                $\theta_\mathrm{LDD}$ (mas) & $38.27 \pm 0.37$ & $39.69 \pm 0.40$ & $38.79 \pm 0.54$ \\
                $\delta_\mathrm{LDD}$ & $0.52 \pm 0.02$ & $0.66 \pm 0.01$ & $0.46 \pm 0.02$ \\
                \hline
        \end{tabular}
\end{table}

\subsection{LDD model with Gaussian hotspots\label{Sect:Nspots}}

Photospheric features on RSG stars are interpreted as the top of convective cells where hot material emerges from the stellar interior. To fit our high angular resolution PIONIER data with such a model, we must use small spot sizes (smaller than 1/10$^\mathrm{}$ of the stellar diameter). Considering the number of small features that can be placed on the photosphere and the many possible locations, classical model fitting is excluded. We rely on comparison of the data with observables computed from an empirical model. The model we use consists of a LDD star on which we add distributions of randomly positioned Gaussian spots of fixed size (described by the full width at half maximum, FWHM). We fit each spectral channel independently.  For each distribution $i$, we can tune the maximum fraction of the visible photosphere that can be occupied by the spots ($f_i$) and the intensity contribution ($I_i$). It is then equally distributed among the individual spots, after a weighting by the limb darkening at the spot's central coordinates:
\begin{equation}
        I_\mathrm{spot}(\mu) = \frac{I_i}{N_{\mathrm{spot}, i}}\mu^{\delta_\mathrm{LDD}}
.\end{equation}

Owing to projection of the sphere onto the plane of the sky, the 
probability of encountering spots near the limb is higher than near the center. 
In the model, spot radial and azimuthal positions are defined by:

\begin{equation}
        r_\mathrm{spot} = \frac{\theta_\mathrm{LDD}}{2}\sqrt{\cos \left(m \times 
        \frac{\pi}{2}\right)}; \\
        \theta_\mathrm{spot} = n \times 2\pi 
,\end{equation}

\noindent where $m$ and $n$ are two independent random variables between 0 and 1. The detailed expression of the complex visibility function for a LDD with a distribution of Gaussian bright spots is (derived from the single spot model of \citeads{2016A&A...588A.130M}):

\begin{equation}
        \begin{array}{l}
V(u, v) = (1-\sum_i I_i)V_\mathrm{LDD}(u, v) + \sum_{i} \sum_{\mathrm{spots}} \left\lbrace I_\mathrm{spot}(u, v) \right. \\\\
\times \exp \left[ -\frac{(2 \pi r_\mathrm{spot} \sigma_i)^2}{2}\right] 
 \exp\left[-2j\pi(u x_\mathrm{spot} + v y_\mathrm{spot})\right] \left. \right\rbrace 
        \end{array}
,\end{equation}

\noindent with $ x_\mathrm{spot} = r_\mathrm{spot}\cos(\theta_\mathrm{spot})$,  $ y_\mathrm{spot} = r_\mathrm{spot}\sin(\theta_\mathrm{spot})$, $ \sigma_i = $~FWHM$_i/(2\sqrt{2\ln(2)})$ and $j^2 = -1$. The expression of $ V_\mathrm{LDD}$ is given by Eq. \ref{Eq:VisLDD}.\\

Since many spot configurations are possible, we consider here the result of 1000 such random models. To have an accurate overview of the observables, we derive the probability of obtaining a given squared visibility or closure phase value. Examples of individual size distributions are given in Appendix \ref{App:Nspot}.

\begin{figure*}
        \resizebox{\hsize}{!}{\includegraphics{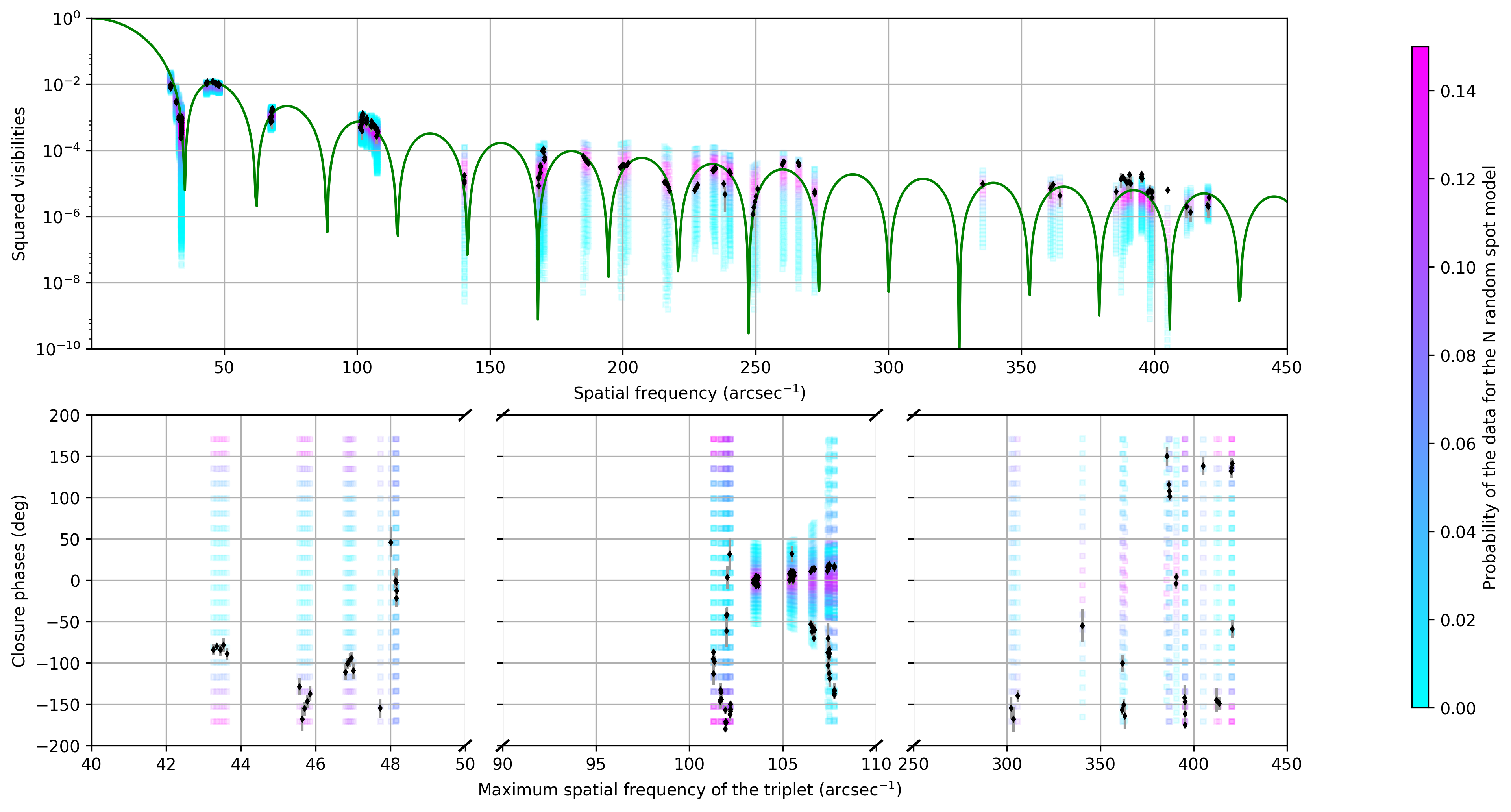}}
        \caption{Interferometric observables measured on Antares by VLTI/PIONIER at 1.61~$\mu$m (black points). The green curve corresponds to a LDD power-law model of 37.89~mas in diameter and a LD exponent of 0.52. The color points represent the probability of the observables computed for 1000 iterations of a LDD power law model of 37.89~mas in diameter and a LD exponent of 0.52 with two distributions of bright Gaussian spots. The first has spots with a FWHM of 17~mas, a filling factor of 0.5 and a contribution of 3\% to the total intensity. The second has spots with a FWHM of 2~mas, a filling factor of 0.4 and a contribution of 10\% to the total intensity. \textit{Top panel:} Squared visibilities. \textit{Bottom panel:} Closure phases. The spatial frequency domain is fragmented to zoom onto each data range. \label{Fig:Nspot_best}}
\end{figure*}

\citetads{2016A&A...588A.130M} showed that the presence of spots directly affects the angular diameter measurement. Therefore, for each distribution we matched with our dataset, we explored a 4 mas range around the best LDD diameter derived in Table \ref{Tab:LDD_by_wlen} with a step of 0.1 mas. In Fig. \ref{Fig:Nspot_best}, we present the model that best matches the data. Its parameters are summarized in Table \ref{Tab:Nspot_best}. As this model exploits both the squared visibilities and the closure phases, for a fair comparison we derived the $\tilde{\chi}^2$ associated to the best fitted LDD alone over the whole spatial frequency range using these two observables ($\tilde{\chi}^2_\mathrm{LDD}$ in Table \ref{Tab:Nspot_best}). We also performed an F-test to determine if the better match of the spot distributions is significant with respect to the null hypothesis (classical LDD). The F parameter is defined by:

\begin{equation}
        F = \frac{\chi^2_\mathrm{LDD} - \chi^2_\mathrm{model}}{N_\mathrm{param,~model} - N_\mathrm{param,~LDD}} \times \frac{N_\mathrm{data} - N_\mathrm{param,~model}}{\chi^2_\mathrm{model}}
        \label{Eq:f_fac}
.\end{equation}

In principle, the test should be performed on the squared visibilities and closure phases separately. Indeed, the closure phase is a function of the triple product that can be derived from the visibilities: a joint test might assume too many degrees of freedom or even be biased. However, this is impossible here because a LDD model fit does not converge on the closure phases alone. We present the F-test on the squared visibilities only in Appendix \ref{App:F_test}. It shows that the spotty model fit is significant in the first and third channels, for squared visibilities taken separately. Here, we present the fit on both observables: the squared visibilities constrain the shape and size of the features while the closure phases deal with the asymmetries and therefore give positional information. We note that in the case of a fibered interferometer like PIONIER, the visibility amplitude is a function of the phase distribution over the pupils, but is independent from the phase delay between the pupils, while the closure phase depends only on the phase delays. Correlations between the two observables may happen at very low fluxes but such sources are too faint to be observable with near infrared interferometry. Therefore we are confident that the joint F-test on closure phases and squared visibilities is reliable on Antares. The critical value for a (2, 5) F distribution\footnote{\url{http://www.itl.nist.gov/div898/handbook/eda/section3/eda3673.htm}} is 5.143. Our F values are much higher than this. Therefore, the lower $\tilde{\chi}^2$ of our distributions of bright Gaussian spots is significant compared to the single LDD model; the better match is not an effect of the increased number of parameters.

\begin{table*}
        \caption{Best matching LDD and Gaussian hotspot distribution model. The $\tilde{\chi}^2$ statistic is computed over 1000 instances of the model and takes into account both the squared visibilities and the closure phases (contrary to the analytical models tested earlier). The $\tilde{\chi}^2_\mathrm{LDD}$ corresponds to the value of the best fitted LDD model over the entire dataset associated to a spectral channel.}
        \label{Tab:Nspot_best}
        \centering
        \begin{tabular}{ll}
                \hline \hline
                \noalign{\smallskip}
                \multicolumn{2}{c}{All channels}\\
                Parameter & Value\\
                \hline
                \noalign{\smallskip}
                FWHM$_\mathrm{dist\ 1}$ & 17~mas \\
            I$_1$ & 3~\% \\
                Filling factor$_\mathrm{dist\ 1}$ & 0.5\\
                FWHM$_\mathrm{dist\ 2}$ & 2~mas \\
                I$_2$ & 10~\% \\
                Filling factor$_\mathrm{dist\ 21}$ & 0.4\\
                \hline
        \end{tabular}
        ~
        \begin{tabular}{llll}
                \hline\hline
                \noalign{\smallskip}
                Parameter & $1.61~\mu$m & $1.66~\mu$m & $1.71~\mu$m \\
                \hline
                \noalign{\smallskip}
                $\theta_\mathrm{LDD}$ (mas) & $37.89 \pm 0.10$ & $38.81 \pm 0.10$ & $38.24 \pm 0.31$\\
                $\delta_\mathrm{LDD}$ (from Table \ref{Tab:LDD_by_wlen})& $0.52 \pm 0.02$ & $0.66 \pm 0.01$ & $0.46 \pm 0.02$ \\
                Mean $\tilde{\chi}^2$ & 82.1 & 109.3 & 113.8\\
                Std. deviation $\tilde{\chi}^2$ & 35.2 & 47.0 & 41.0 \\
                Min $\tilde{\chi}^2$ & 28.1 & 37.6 & 43.8 \\
                $\tilde{\chi}^2_\mathrm{LDD}$ & 622 & 1611 & 1447 \\
                F & 2818 & 5859 & 4495 \\
                \hline
        \end{tabular}
\end{table*}

However, our best match model should not be taken as the characteristics of the actual feature distribution. Indeed, the actual features are probably not symmetric Gaussians and probably have a continuous size distribution. Our model expresses a possible scenario to reproduce the observed squared visibilities and closure phases. In particular, it offers an explanation for the enhanced power of the high-frequency squared visibilities compared to the LDD, and for the spread of the squared visibilities depending on the PA. 

Among the different combinations we explored, two fundamental conclusions arise. Firstly, the cloud-like shape of the squared visibilities and closure phases at high spatial frequencies (higher than 300~arcsec$^{-1}$) can only be reproduced by introducing non-resolved spots (e.g., with a FWHM no greater than 2~mas). Secondly, the introduction of a distribution of spots with any FWHM leads to modifications of the shape of the first and second lobes of the squared visibility function (see Appendix \ref{App:Nspot}). Therefore, the presence of bright features on the stellar surface has direct consequences on the low-spatial-frequency range historically used to determine the angular diameter of the star and its LD. 

\subsection{Limb darkening}

In Table \ref{Tab:LDD_rsg}, we present LD measurements on several RSG stars and compare them to our results on Antares at 1.61~$\mu$m. We considered only power-law models.

\begin{table}
        \caption{LD power law exponent for different RSGs}
        \label{Tab:LDD_rsg}
        \centering
        \begin{tabular}{lll}
                \hline \hline
                \noalign{\smallskip}
                Name & LD parameter & Ref. \\
                \hline
                \noalign{\smallskip}
                Antares & $ 0.52 \pm 0.02$ & (1)\\
                Betelgeuse & $0.23 \pm 0.09$ & (2)\\    
                \object{$\alpha$ Her} & $0.394 \pm 0.180$ & (3)\\
                \object{T Per} & $0.32 \pm 0.2$  & (4)\\
                \object{RS Per} & $0.34 \pm 0.2$  & (4)\\
                \hline
        \end{tabular}
        \tablebib{
                (1)~This work;
                (2)~\citetads{2016A&A...588A.130M};     
                (3)~\citetads{2004A&A...418..675P};
                (4)~\citetads{2014ApJ...785...46B}
        }
\end{table}

The value of the LD parameter is higher than the previous measurement in the K band continuum by O13. Although we have analyzed our dataset differently, we would like to state that even by considering the whole spatial frequency range, we reach a LD parameter of $\sim 0.4$; still higher than theirs. This makes it also higher than similar observations on other RSGs. However, as we have only one PA direction in the second lobe to constrain the LD measurement, we cannot exlude that this value is biased by photospheric features.

\section{Numerical approach: radiative hydrodynamics simulation\label{Sect:RHD_Sim}}

To go beyond analytical models, we now turn to numerical convective simulations based on 3D radiative hydrodynamics computations. On Betelgeuse, 
C10b and \citetads{2014A&A...572A..17M} managed to reproduce the measured 
squared visibilities with such numerical models. However, \citetads{2016A&A...588A.130M} 
did not reproduce both the squared visibilities and the closure phases of their H band PIONIER observations of the same star. They suggested 
that the convective activity was disturbed by the presence of a large hot spot 
identified as the top of a huge convective cell. 

To fit our PIONIER data of Antares, we used two simulations obtained with the CO$^5$BOLD code (COnservative COde for the COmputation of
COmpressible COnvection in a BOx of L Dimensions, L = 2, 3, 
\citeads{2012JCoPh.231..919F}), with stellar parameters close to those derived on Antares (O13). The characteristics of these numerical models 
are given in Table \ref{Tab:Charac_Simu}. We note that rotation is not yet 
implemented in these models. Computation of RHD models is complex and very 
demanding of computer resources, and it has not yet been possible to prepare a 
suite of models specifically tuned to Antares. For now, we choose the best 
available models, and leave for the future custom model generation and 
iteration for each star studied.

\begin{table*}
        \caption{Characteristics of the RHD simulations used to analyze our VLTI/PIONIER data. (See C11a for more details).}
        \label{Tab:Charac_Simu}
        \centering
        \begin{small}
        \begin{tabular}{cccccccc}
                \hline\hline
                \noalign{\smallskip}
                Model & L & T$_\mathrm{eff}$ & R$_\star$ & $\log g$ & Grid & Grid \\
                & (L$_\odot$) & (K) & (R$_\odot$) & & (N & res. \\
                & & & & & points) & [$R_\odot$] \\
                \hline
                \noalign{\smallskip}
                st35gm03n07 & $91\,932 \pm 1400$ & $3487 \pm 12$ & $830.0 \pm 2.0$ & $-0.335 \pm 0.002$ & $235^3$ & 8.6 \\
                st35gm03n13 & $89\,477 \pm 857$ & $3430 \pm 8$ & $846.0 \pm 1.1$ & 
                $-0.354 \pm 0.001$ & $235^3$ & 8.6 \\
                \hline
        \end{tabular}
        \end{small}
\end{table*}

We checked that we do not reach the numerical limit induced by the spatial gridding of the simulations. According to \citetads[][Eq.3]{2009A&A...506.1351C}, artifacts will affect the derived visibilities for spatial frequencies higher than 0.03~$R_\odot^{-1}$. Using the equation:
\begin{equation}
\nu [\mathrm{arcsec}^{-1}] = \nu [\mathrm{R}_\odot^{-1}] \cdot d[\mathrm{pc}] \cdot 214.9
,\end{equation}
we convert this to 1250~arcsec$^{-1}$.
We note that the distance in parsec is not the actual distance between the solar system and Antares but the distance required for the stellar model to have the same apparent angular size as Antares. Therefore, our VLTI/PIONIER data, although with a very high resolution, are still below the frequency of expected artifacts in the simulations.\\

For each simulation, hundreds of temporal snapshots were computed; each of them is a realization of the convective pattern of the star. Using the 3D
pure-LTE (local thermodynamical equilibrium) radiative transfer code Optim3D \citepads{2009A&A...506.1351C}, intensity images are computed in the three spectral channels of our PIONIER observations. As Antares may have any orientation on the plane of the sky relative to the simulation, we rotated each image around its center. We used 36 angle positions between 0$^\circ$ and 180$^\circ$. The distance of Antares was taken into account by scaling the angular diameters to the value we derived from the LD power law. Interferometric observables were computed using a Fast Fourier Transform algorithm.

Contrary to previous matches of interferometric data with these simulations (C10b, \citeads{2010A&A...511A..51C}, \citeads{2014A&A...572A..17M,2016A&A...588A.130M} or \citeads{2015A&A...575A..50A}) , we did not seek the best matching snapshot and rotation angle. Instead, we considered the $\tilde{\chi}^2$ associated to the whole grid of temporal snapshots and rotation angles for both simulations. The characteristics of these $\tilde{\chi}^2$ distributions are given in Table \ref{Tab:RHD_sim_fit}. The closure phases are mostly constrained by positional information of the inhomogeneities. Therefore, we only considered the squared visibilities that give mainly information about the number and size of the stellar features.

\begin{table}
        \caption{Comparison of the RHD simulations with the PIONIER data of Antarès. The $\tilde{\chi}^2$ is computed over the entire squared visibility dataset.}
        \label{Tab:RHD_sim_fit}
        \centering
        \begin{tabular}{llll}
                \hline \hline
                \noalign{\smallskip}
                 & \multicolumn{3}{c}{$\tilde{\chi}^2$}\\
                Simulation & mean & min. & std. dev. \\
                \hline
                \noalign{\smallskip}
                st35gm03n07 & 1838 & 356 & 1178\\
                st35gm03n13 & 871 & 252 & 326 \\
                \hline
        \end{tabular}
\end{table}

The simulation st35gm03n13 gives a better match to the observed squared visibilities. It has a non-gray opacity 
approximation (we refer to C11a for the details about this physical approximation). 
This causes an intensified heat exchange of a fluid element with its 
environment, reducing the temperature/density fluctuations (Fig. 5 in C11a). 
Less intense fluctuations reduce the surface intensity contrast of nearby areas 
and, eventually, interferometric observables. On the contrary, the st35g03n07 
simulation has a gray opacity approximation.

\begin{figure}
        \centering
        \resizebox{\hsize}{!}{\includegraphics{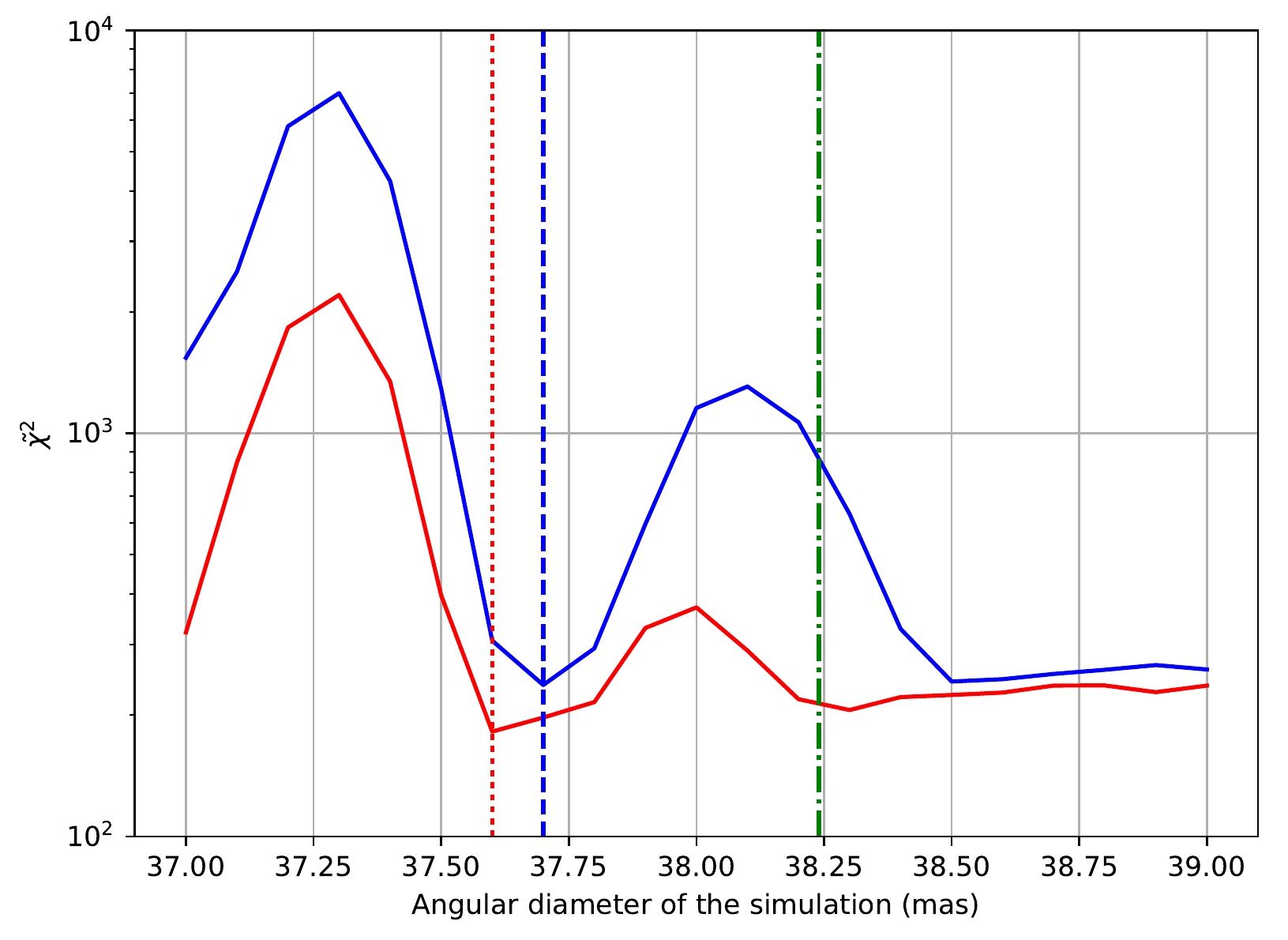}}
        \caption{Mean (blue) and minimum (red) $\tilde{\chi}^2$ computed over the whole rotation angles of the best matching snapshot of simulation st35gm03n13, as a function of the simulation angular diameter at 1.61~$\mu$m. The red vertical dotted line corresponds to the diameter associated to the minimum of the continuous red curve. The blue vertical dashed line corresponds to the diameter associated to the minimum of the continuous blue curve. The green vertical dash-dotted line corresponds to the best LDD angular diameter obtained in Sect. \ref{Sect:UD_LDD}. The $\tilde{\chi}^2$ is computed over the entire squared visibility dataset. \label{Fig:simu_diameter}}
\end{figure}

We saw in Sect. \ref{Sect:Analytical} that the angular diameter determination is strongly sensitive to the presence of stellar features as they may
directly affect the squared visibilities in the first lobe. Thus, for the best matching temporal snapshot of the best simulation, we generated squared visibilities matching our PIONIER observations for a stellar model between 37.0 and 39.0~mas with a step of 0.1~mas, and we kept the 36 rotation angles. The mean and minimum $\tilde{\chi}^2$ are represented on Fig. \ref{Fig:simu_diameter}. It appears that the $\tilde{\chi}^2$ value strongly depends on the fixed angular diameter of the simulation. In particular, the simulations computed with the best LDD diameter ($38.24 \pm 0.37$~mas) derived in Sect. \ref{Sect:UD_LDD} do not give the best result. The minimum $\tilde{\chi}^2$ of 182.2 is reached for an angular diameter of $37.60 \pm 0.08$~mas at 1.61~$\mu$m. This smaller size between the 3D simulations and the classical angular diameter fit of interferometric data is consistent with previous results from \citetads{2010A&A...524A..93C} that noted a similar discrepancy with K giant observations. The main origin would be the presence of bright inhomogeneities that force a smaller angular diameter in order to keep the same integrated luminosity over the stellar disk.

However, this $\tilde{\chi}^2$ , computed over the squared visibilities only, remains worse than what is achievable with the power law LDD alone ($\sim 25$ for the whole squared visibility data in Sect. \ref{Sect:UD_LDD}) or the random distribution of Gaussian spots ($\sim 28 - 44$ in Sect. \ref{Sect:Nspots} for both closure phases and visibilities, that should be in principle more difficult to fit). This is also what was obtained on the RSG Betelgeuse \citepads{2016A&A...588A.130M}. This comes as a surprise as previous interferometric observations in the H (C10b) and K bands \citepads{2014A&A...572A..17M} were well matched by 3D RHD simulations. It is not within the scope of this paper to study the quality of the match between observations and simulations over time for RSG stars but we would like to stress the importance of continuous monitoring of these stars over time with various interferometric instruments. Their convective patterns do not appear to be always reproducible by current state-of-the-art RHD simulations.

\section{Conclusions}

\subsection{The angular diameter, limb-darkening, and photospheric features}

We present highly resolved interferometric observations of the nearby RSG 
Antares using the three available AT configurations of the VLTI with the 
PIONIER instrument. These observations have allowed us to study the angular diameter 
and limb-darkening of the star. 

The deviations from the uniform disk and limb-darkened disk models depend on the position angle of the measurements and affect the derived values for the angular diameter while not affecting the limb-darkening. In Sect. \ref{Sect:UD_LDD} we determined that the photospheric features could affect the angular diameter measurement even if one only considers low- spatial-frequencies data (first and second lobe squared visibility). The effect of these inhomogeneities can only be traced if several PA are probed by the interferometric observations. In particular, RSG observations covering a single direction of the ($u, v$) plane cannot reveal such bias in the angular diameter. Consequently, the angular diameter derived from the averaged angular diameter over several PA can only be a rough approximation. A proper estimation of the angular diameter needs to use a model that generates photospheric features consistent over the whole squared visibility and closure phase data. When doing so by adding a single bright Gaussian spot model over a LDD (on Betelgeuse, \citeads{2016A&A...588A.130M}), or a distribution of Gaussian spots or even a 3D RHD simulation (on Antares, present work), the derived angular diameter is smaller than with a classical featureless model.

\subsection{The convective signature}

We also showed (subject to the assumptions of Sect. \ref{Sect:Nspots} on the joint F-test on both squared visibilities and closure phases) that the departure from the disk model at intermediate and high 
spatial frequency was compatible with a LDD model combined with distributions of Gaussian spots. Additionally, it is also qualitatively compatible with convective RHD simulations. This means 
that our data detected small convective cells. Our dataset does not reach the 
numerical threshold of simulations beyond which artifacts can appear.

\citetads{1975ApJ...195..137S} made several predictions concerning the sizes and number of photospheric features for RSGs. Using physical models to derive the quantities characterizing the subphotospheric convection of a RSG, he derived a characteristic size of $0.14~R_\star$ (within the scenario of convective cells of the maximum possible physical scale). In the case of Antares, this would correspond to elements with a characteristic size of 2.7~mas. We see that even his largest estimation corresponds to the smaller features we are able to resolve (Sect.\ref{Sect:Nspots}).

C10b and \citetads{2014A&A...572A..17M,2016A&A...588A.130M} studied the convective surface of the prototypical RSG Betelgeuse. The various datasets of C10b and the VLTI/AMBER observation of \citetads{2014A&A...572A..17M} were rather well matched by convective simulations. However, the four epochs of the VLTI/PIONIER observations of \citetads{2016A&A...588A.130M} cannot be reproduced by 3D RHD simulations due to the presence of a bright surface feature on the stellar disk.

For our PIONIER observations of Antares, the signal is not well reproduced either but we determine that the match is strongly affected by the angular diameter imposed on the simulation. It appears that the reproduction of photospheric features on RSG by 3D RHD simulations still requires improvement of the physical recipes of the model. Temporal monitoring at different wavelengths of several RSG stars would help constrain those missing ingredients by providing more examples of convective patterns.

\begin{acknowledgements}
        We are grateful to the Paranal Observatory team for the successful execution of the observations.
        This research received the support of PHASE, the high angular resolution partnership between ONERA, Observatoire de Paris, CNRS and University Denis Diderot Paris 7.
        We acknowledge financial support from the ``Programme National de Physique Stellaire" (PNPS) of CNRS/INSU, France.
        The authors would like to thank Alain Chelli for his useful advice on the handling of closure phases and squared visibilities from a statistical point of view.
        We used the SIMBAD and VIZIER databases at the CDS, Strasbourg (France)\footnote{Available at \url{http://cdsweb.u-strasbg.fr/}}, and NASA's Astrophysics Data System Bibliographic Services.
        This research has made use of Jean-Marie Mariotti Center's \texttt{Aspro}\footnote{Available at \url{http://www.jmmc.fr/aspro}} service, of the \texttt{LITpro}\footnote{Available at \url{http://www.jmmc.fr/litpro}} software (co-developped by CRAL, LAOG and FIZEAU) and of the \texttt{SearchCal} service\footnote{Available at \url{http://www.jmmc.fr/searchcal}} (co-developped by FIZEAU and LAOG/IPAG).
        This research made use of IPython \citep{PER-GRA:2007} and 
        Astropy\footnote{Available at \url{http://www.astropy.org/}}, a 
        community-developed core Python package for Astronomy 
        \citepads{2013A&A...558A..33A}.
\end{acknowledgements}

\bibliographystyle{aa}
\bibliography{./biblio}

\begin{appendix} 
        \section{Log of the VLTI/PIONIER observations}
        
        Our VLTI/PIONIER observations of Antares and its calibrators are given for each array configuration in Table \ref{Tab:ObsLog}.  
        
        \begin{table}[!ht]
                \caption{Log of the PIONIER observations of Antares and its calibrators}
                \label{Tab:ObsLog}
                \centering
                \begin{tabular}{llll}
                        \hline \hline
                        \noalign{\smallskip}
                        UT & & Star & Configuration\\
                        \hline
                        \noalign{\smallskip}
                        2014-Apr-24 & 04:17 & HR 5969 & A1-B2-C1-D0\\
                        & 04:41 & Antares & A1-B2-C1-D0\\
                        & 05:03 & HR 6145 & A1-B2-C1-D0\\
                        & 05:19 & Antares & A1-B2-C1-D0\\
                        & 05:38 & HD 148643 & A1-B2-C1-D0\\
                        & 05:51 & Antares & A1-B2-C1-D0\\
                        & 06:08 & HR 5969 & A1-B2-C1-D0\\
                        & 06:22 & Antares & A1-B2-C1-D0\\
                        & 06:34 & HR 6145 & A1-B2-C1-D0\\
                        & 06:47 & Antares & A1-B2-C1-D0\\
                        & 07:04 & $\psi$ Oph & A1-B2-C1-D0\\ 
                        & 07:16 & Antares & A1-B2-C1-D0\\
                        & 07:43 & Antares & A1-B2-C1-D0\\ 
                        & 07:44 & HR 5969 & A1-B2-C1-D0\\ 
                        & 07:57 & Antares & A1-B2-C1-D0\\
                        & 08:15 & $\psi$ Oph & A1-B2-C1-D0\\
                        & 08:30 & Antares & A1-B2-C1-D0\\   
                        & 08:43 & HR 5969 & A1-B2-C1-D0\\
                        & 08:56 & Antares & A1-B2-C1-D0\\
                        & 09:09 & $\psi$ Oph & A1-B2-C1-D0\\
                        & 09:26 & Antares & A1-B2-C1-D0\\
                        & 09:39 & HR 5969 & A1-B2-C1-D0\\ 
                        & 09:52 & Antares & A1-B2-C1-D0\\
                        & 10:05 & $\psi$ Oph & A1-B2-C1-D0\\
                        & 10:18 & Antares & A1-B2-C1-D0\\
                        2014-Apr-29 & 04:44 & HR 5969 & D0-H0-G1-I1\\
                        & 05:32 & HD 148643 & D0-H0-G1-I1\\
                        & 06:18 & Antares & D0-H0-G1-I1\\
                        2014-May-04 & 07:41 & HD 148643 & A1-G1-K0-J3\\
                        & 08:03 & Antares & A1-G1-K0-J3\\
                        & 08:26 & Antares & A1-G1-K0-J3\\
                        & 08:37 & HR 5969 & A1-G1-K0-J3\\
                        & 09:09 & HD 148643 & A1-G1-K0-J3\\
                        & 09:27 & Antares & A1-G1-K0-J3\\
                        & 09:39 & HR 6145 & A1-G1-K0-J3\\
                        & 10:04 & Antares & A1-G1-K0-J3\\
                        2014-May-7 & 06:16 & HD 142407 & A1-G1-K0-J3\\
                        & 06:36 & Antares & A1-G1-K0-J3\\
                        & 07:00 & HD 143900 & A1-G1-K0-J3\\
                        & 07:17 & Antares & A1-G1-K0-J3\\
                        & 07:33 & HR 6145 & A1-G1-K0-J3\\
                        & 07:46 & Antares & A1-G1-K0-J3\\
                        & 08:04 & HD 142407 & A1-G1-K0-J3\\
                        & 08:21 & Antares & A1-G1-K0-J3\\
                        & 08:44 & HD 143900 & A1-G1-K0-J3\\
                        & 09:05 & Antares & A1-G1-K0-J3\\
                        & 09:30 & HR 6145 & A1-G1-K0-J3\\
                        & 09:45 & Antares & A1-G1-K0-J3\\
                        \hline
                \end{tabular}
        \end{table}
        
        \section{Example of random Gaussian spot distribution in a limb-darkened disk\label{App:Nspot}}
        
\begin{figure*}[ht!]
                \resizebox{\hsize}{!}{\includegraphics{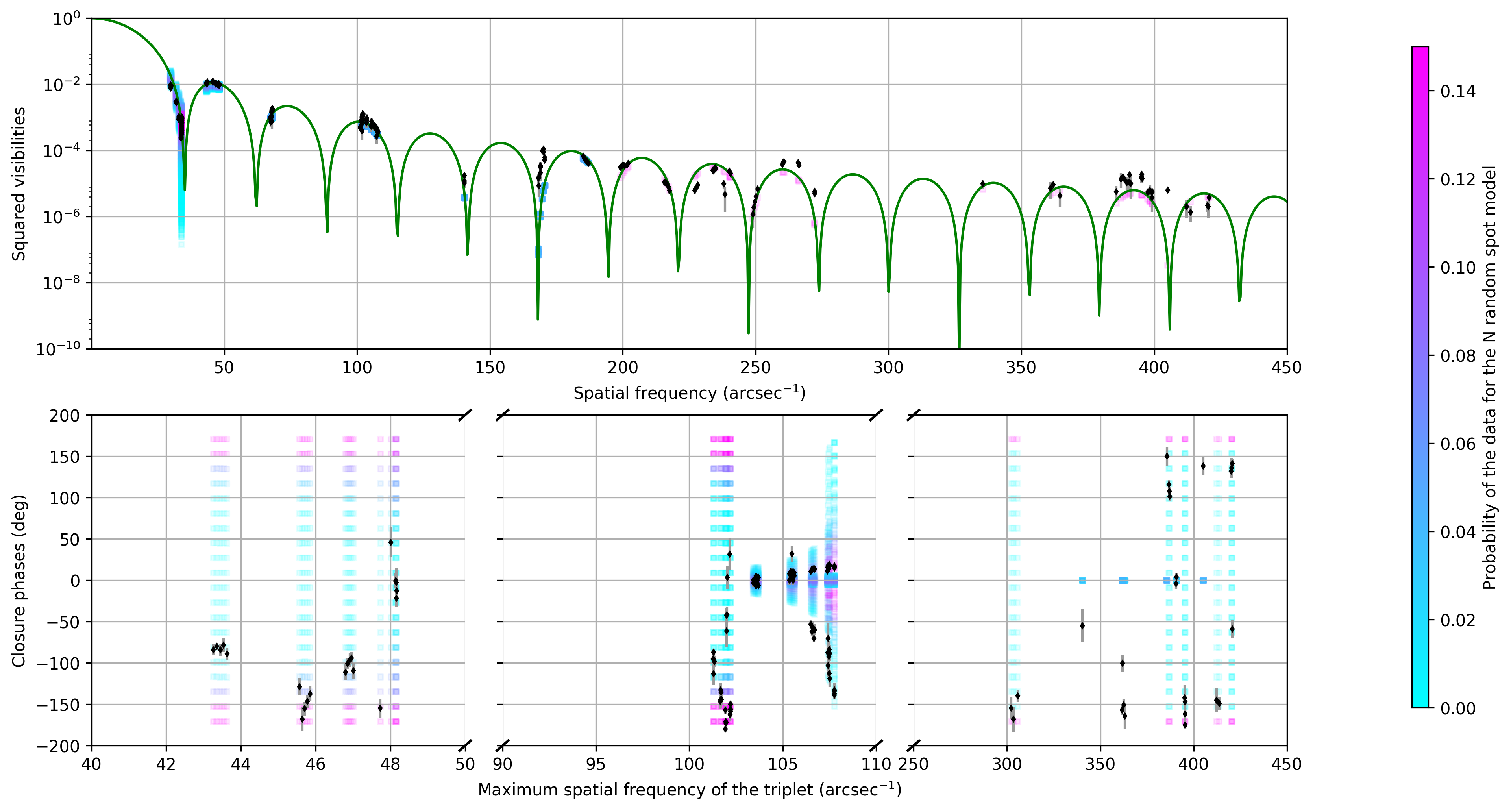}}
                \caption{Interferometric observables measured on Antares by VLTI/PIONIER at 1.61~$\mu$m (black points). The green curve corresponds to a LDD power-law model of 37.89~mas in diameter and a LD exponent of 0.52. The color points represent the probability of the observables recorded on 1000 iterations of a LDD power law model of 37.89~mas in diameter and a LD exponent of 0.52 with a distribution of bright Gaussian spots with a FWHM of 17~mas, a filling factor of 0.5 and a contribution of 10\% to the total intensity. \textit{Top panel:} Squared visibilities. \textit{Bottom panel:} Closure phases. The spatial frequency domain is fragmented to zoom onto each range. \label{Fig.Nspot_17mas}}
        \end{figure*}

\begin{figure*}[ht!]
                \resizebox{\hsize}{!}{\includegraphics{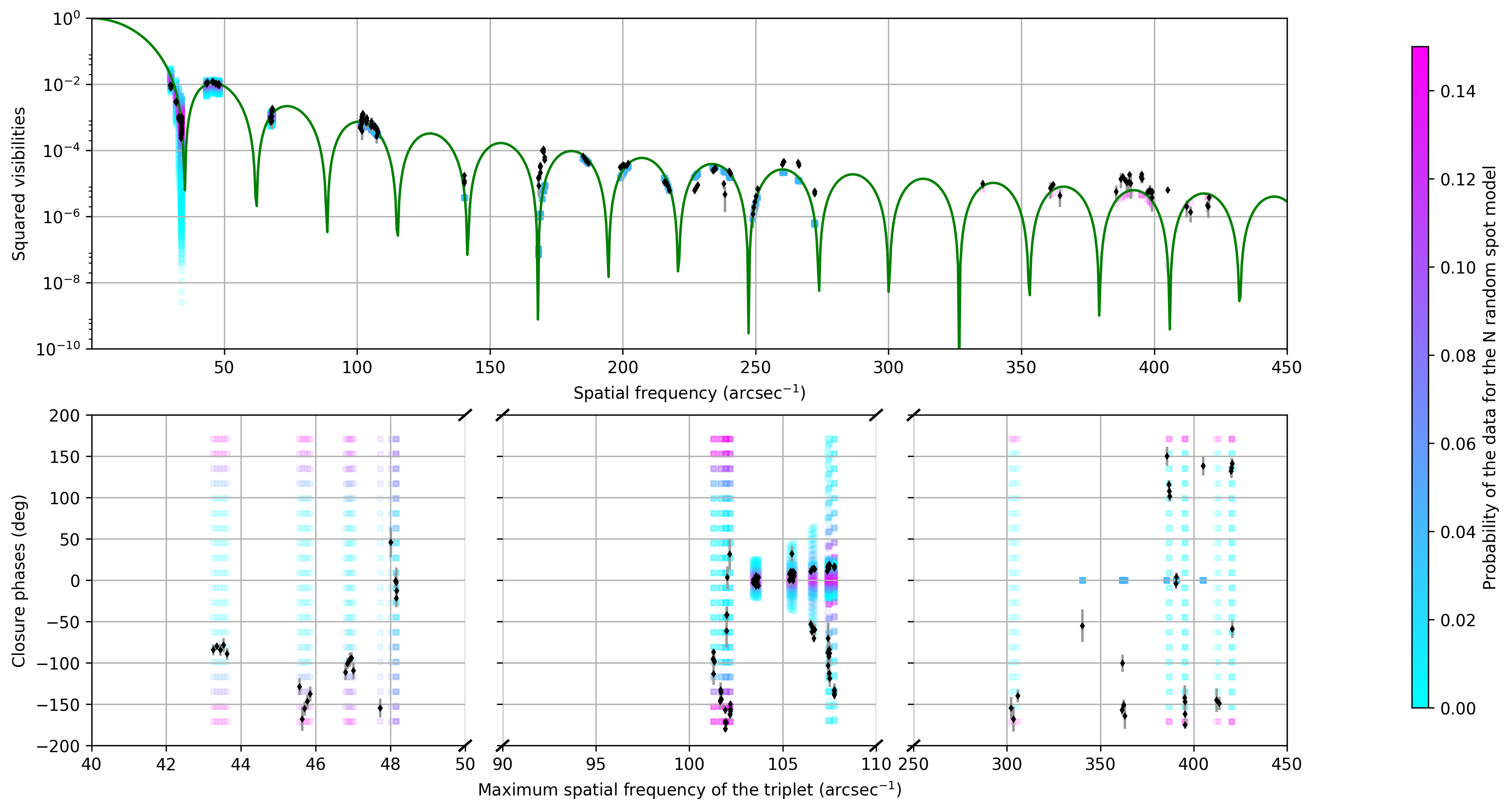}}
                \caption{Identical to Fig. \ref{Fig.Nspot_17mas} with a distribution of spots with a FWHM of 12~mas, a filling factor of 0.5 and a contribution of 10\% to the total intensity. \label{Fig.Nspot_12mas}}
        \end{figure*}

\begin{figure*}[ht!]
                \resizebox{\hsize}{!}{\includegraphics{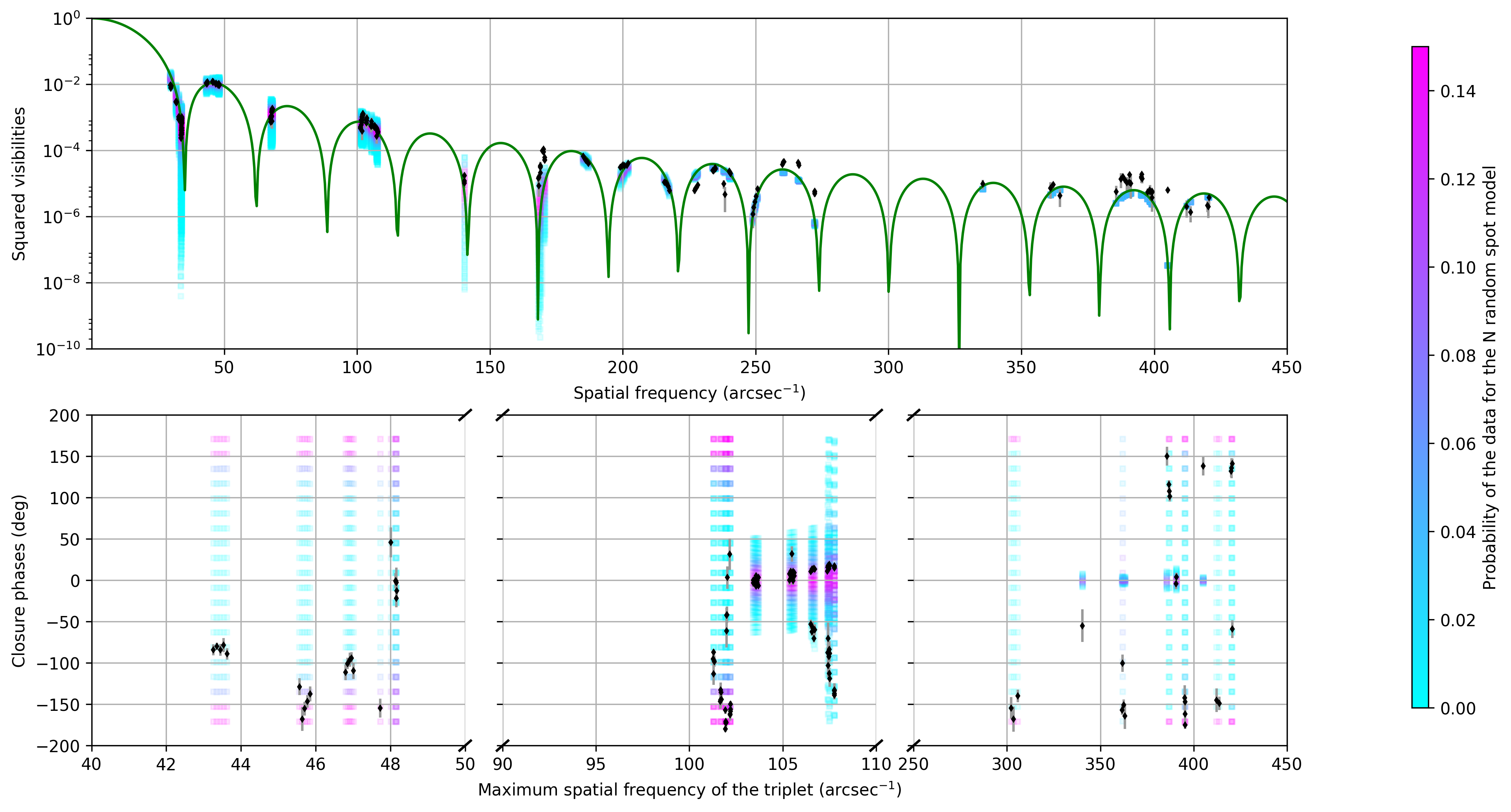}}
                \caption{Identical to Fig. \ref{Fig.Nspot_17mas} with a distribution of spots with a FWHM of 5~mas, a filling factor of 0.4 and a contribution of 10\% to the total intensity. \label{Fig.Nspot_5mas}}
        \end{figure*}

\begin{figure*}[ht!]
                \resizebox{\hsize}{!}{\includegraphics{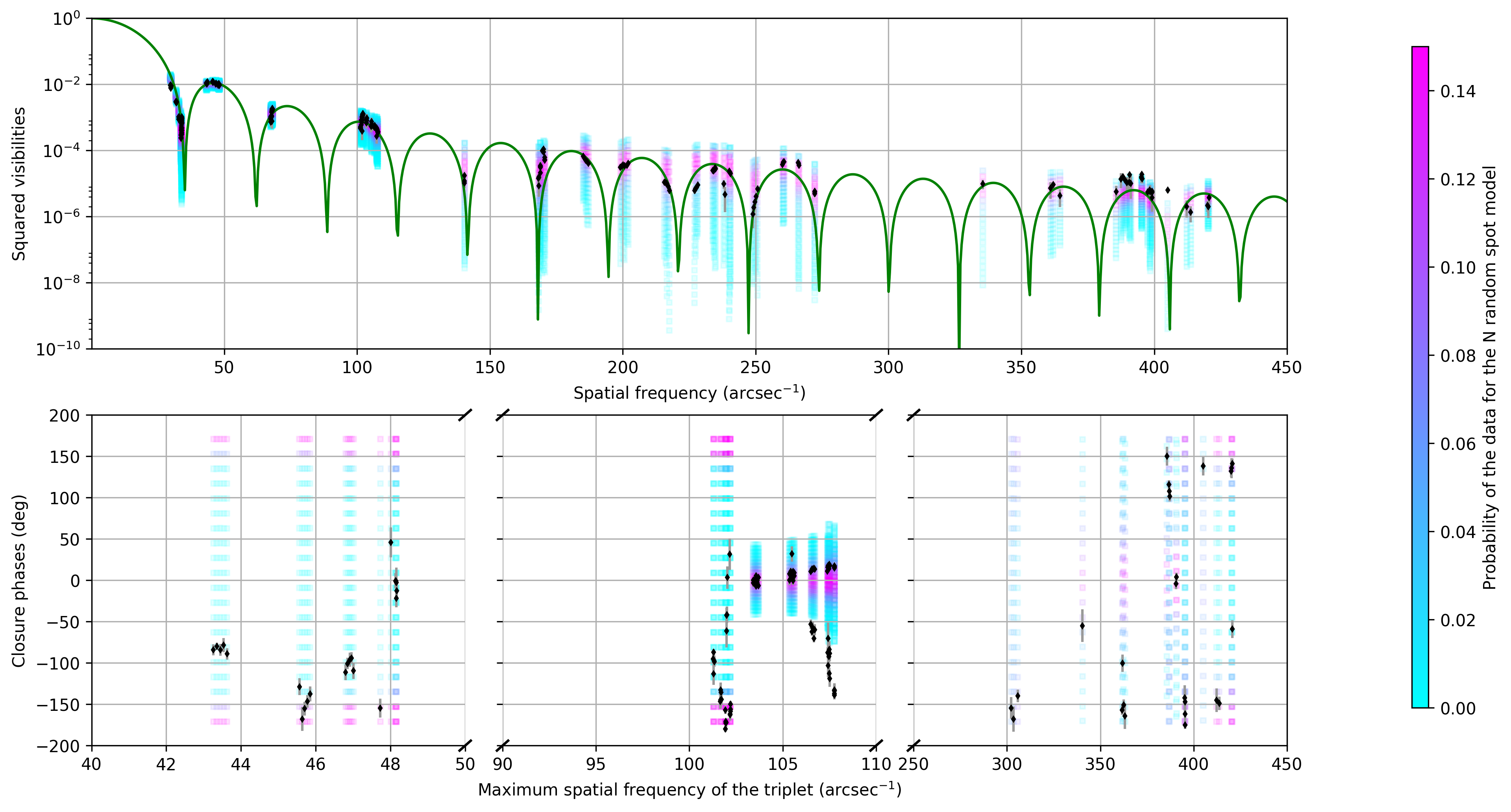}}
                \caption{Identical to Fig. \ref{Fig.Nspot_17mas} with a distribution of spots with a FWHM of 2~mas, a filling factor of 0.3 and a contribution of 10\% to the total intensity. \label{Fig.Nspot_2mas}}
        \end{figure*}

        To obtain more details on the model used in the following examples, we refer the reader to Sect. \ref{Sect:Nspots}. The example are presented in Figs. \ref{Fig.Nspot_17mas}, \ref{Fig.Nspot_12mas}, \ref{Fig.Nspot_5mas} and \ref{Fig.Nspot_2mas}.
        
        The decreasing size of the bright Gaussian spots creates more dispersion of the squared visibilities at higher and higher spatial frequencies and also increases the closure phase signal (deviation from 0$^\circ$ or 180$^\circ$).

        \section{F-test to determine the significance of the limb-darkening disk and Gaussian spot distribution model fit on the squared visibilities \label{App:F_test}}
        
        We performed an F-test to determine if the better match of the spot distribution model is significant with respect to the null hypothesis (classical LDD). The F parameter expression is given in Eq. \ref{Eq:f_fac}. We tried to perform the test separately on the squared visibilities and closure phases. However, it was impossible to make the LDD model (without spots) fit to converge with the closure phases only. Therefore, we only present the F-test on the squared visibilities. The critical value for a (2, 5) F distribution\footnote{\url{http://www.itl.nist.gov/div898/handbook/eda/section3/eda3673.htm}} is 5.143. Therefore, from the values of Table \ref{Tab:Ftest}, our model fitting with bright spots is significant only in the first and third spectral channels of our VLTI/PIONIER squared visibilities. This could mean that at 1.66~$\mu$m the features are less prominent but this is without taking into account the closure phase signal. In Sect. \ref{Sect:Nspots}, we present the fit on both the squared visibilities and closure phases with a F factor much higher, meaning that the closure phases bring information about the asymmetries that are reproduced by our model. 
        
        \begin{table*}[ht!]
                \caption{Result of the F-test performed on the LDD and random Gaussian spot distribution models on the squared visibilities only. The LDD model fit did not converged when using only the closure phases.\label{Tab:Ftest}}
                \centering
                \begin{tabular}{lllll}
                \hline\hline
                \noalign{\smallskip}
                Model & Parameter & $1.61~\mu$m & $1.66~\mu$m & $1.71~\mu$m \\
                \hline
                \noalign{\smallskip}
                & $\theta_\mathrm{LDD}$ (mas) & $37.84 \pm 0.31$ & $38.81 \pm 0.31$ & $38.24 \pm 0.10$\\
                
                \multirow{2}{*}{LDD with} & $\delta_\mathrm{LDD}$ (from Table \ref{Tab:LDD_by_wlen})& $0.52 \pm 0.02$ & $0.66 \pm 0.01$ & $0.46 \pm 0.02$ \\
                 
                \multirow{2.5}{*}{spot dist.} & Mean $\tilde{\chi}^2$ & 71 & 58 & 51\\
                & Std. deviation $\tilde{\chi}^2$ & 41 & 29 & 29 \\
                & Min $\tilde{\chi}^2$ & 16 & 23 & 17 \\
                \noalign{\smallskip}
                \multirow{1.5}{*}{LDD.} & $\theta_\mathrm{LDD}$ (mas) & $37.81 \pm 0.09$ & $38.14 \pm 0.09$ & $37.97 \pm 0.09$\\
                \multirow{1.5}{*}{alone} & $\delta_\mathrm{LDD}$ (mas) & $0.45 \pm 0.02$ & $0.40 \pm 0.01$ & $0.34 \pm 0.02$\\
                & $\tilde{\chi}^2_\mathrm{LDD}$ & 21 & 17 & 19 \\
                \noalign{\smallskip}
                & F & 31 & -21 & 11 \\
                \hline
        \end{tabular}
        \end{table*}
        
\end{appendix}

\end{document}